\documentclass{copernicus2}

\newcommand{\aap}{    {Astron. Astrophys.}}

\newcommand{\apj}{    {Astrophys. J.}}
\newcommand{\apjl}{   {Astrophys. J. Lett.}}

\newcommand{\jgr}{    {J. Geophys. Res.}}
\newcommand{\mnras}{  {Mon. Not. Roy. Astron. Soc.}}

\newcommand{\planss}{ {Planet. Space Sci.}}
\newcommand{\pre}{    {Phys. Rev. E}}
\newcommand{\solphys}{{Solar Phys.}}

\newcommand{\ssr}{    {Space Sci. Rev.}}

\begin{document}


\title{MHD flows at astropauses and in astrotails}

\author[1]{D. H. Nickeler}
\author[2]{T. Wiegelmann}
\author[1]{M. Karlick\'y}
\author[1]{M. Kraus}

\affil[1]{Astronomical Institute, AV \v{C}R, Fri\v{c}ova 298,
25165 Ond\v{r}ejov, Czech Republic}
\affil[2]{Max-Planck-Institut f\"{u}r Sonnensystemforschung, 
Justus-von-Liebig-Weg 3, 37077 G\"{o}ttingen}


\runningtitle{MHD flows at astropauses and in astrotails}

\runningauthor{Nickeler et al.}

\correspondence{D. H. Nickeler\\ (dieter.nickeler@asu.cas.cz)}

\received{}
\pubdiscuss{} 
\revised{}
\accepted{}
\published{}


\firstpage{1}

\maketitle  

\noindent {\bf ABSTRACT}\\

\noindent 
The geometrical shapes and the physical properties of stellar wind -- interstellar 
medium interaction regions form an important stage for studying stellar winds and their 
embedded magnetic fields as well as cosmic ray modulation. 
Our goal is to provide a proper representation and classification of counter-flow configurations 
and counter-flow interfaces in the frame of fluid theory. 
In addition we calculate flows and large-scale electromagnetic fields based on which 
the large-scale dynamics and its role as possible background for particle acceleration, e.g.
in the form of anomalous cosmic rays, can be studied.
We find that for the definition of the boundaries, which are determining the astropause shape,
the number and location of magnetic null points and stagnation points is essential. 
Multiple separatrices can exist, forming a highly complex environment for the interstellar 
and stellar plasma. Furthermore, the formation of extended tail structures occur naturally,
and their stretched field and streamlines provide surroundings and mechanisms for the 
acceleration of particles by field-aligned electric fields. 


\introduction  

When stars move through the interstellar medium (ISM), the material released via their winds 
collides and interacts with the ISM. This interaction produces several detectable
structures, such as stellar wind bow shocks when stars move with supersonic speeds relativ 
to the ISM. Furthermore, a termination shock can form, where the supersonic stellar wind 
slows down to subsonic speed, and in between this termination shock and the outer bow shock
a contact surface forms, separating the subsonic ISM flow from the subsonic stellar wind flow.
This contact surface is called the astropause and has at least one stagnation point at which
both flows, the stellar wind and the ISM material stop and diverge.


In downwind direction, a tail like structure can
form, which is the astrotail. The tail is not only proposed by the result of simulations, but also 
by the fact that stretched field or streamlines minimize the corresponding tension forces, so 
that the configuration is able to approach an equilibrium state. Such bow shocks 
and astrotails have been directly observed, e.g., around asymptotic giant branch stars 
\citep[e.g.,][]{2008ApJ...687L..33U, 2010ApJ...711L..53S}, and have been proposed to exist also around the Sun \citep[e.g.,][]{2014ApJ...782...25S}.

Typically, a magnetic field is embedded in the ISM. For stars with a strong magnetic field, the 
wind is magnetized as well. Hence, different null points can appear, one of the flow and one of 
the magnetic field, which are not necessarily at the same location, unless the magnetic field is 
frozen-in \citep[see,][]{2008ASTRA...4....7N}. The formation of a 
stagnation point of the flow or of a magnetic neutral point is crucial for the understanding how 
an interface in the form of an astropause forms between the very local ISM and a stellar wind. 
The best object to study counterflow configurations observationally is the heliosphere, where 
spacecrafts such as {\it Voyager 1 \& 2} and {\it IBEX} perform in situ measurements of the 
plasma parameters \citep[e.g.,][]{2013Sci...341..147B, 
2013ApJ...771...77M, 2014A&A...561A..74F}. However, an 
interpretation of these observations is not always straight forward. While strong indications 
for the crossing of the termination shock of {\it Voyager 1} exist, it is yet unclear and 
contradictory whether the heliopause region was already left \citep{2013Sci...341..147B, 2014ApJ...784..146B, 
2013ApJ...776...79F, 2013Sci...341.1489G}. 
 
The problematics of computing counterflow configurations have been attacked from different 
viewpoints, such as hydrodynamics \citep[see, e.g.,][]{1983A&A...118...57F}, 
kinematical magnetohydrodynamics (MHD) \citep[see, e.g.,][]{1990JGR....95.6403S,
1991A&A...250..556N, 1993JGR....9815169N, 1995JGR...100.3463N}, and self-consistent 
MHD \citep[see, e.g.,][]{1982MNRAS.202..735N, 1983MNRAS.205..839F, 2001ohnf.conf...57N,
2005AdSpR..35.2067N, 2006AdSpR..37.1292N, 2006A&A...454..797N, 2006ASTRA...2...63N, 
2008ASTRA...4....7N}, focusing on different aspects and regions within the 
astrosphere. We are aware that fluid approaches like MHD are strictly valid 
only for collisional plasmas, but are frequently applied to collisionless 
configurations like magnetospheres, coronae or astrospheres. The alternative 
approach, kinetic theory, is not applicable because of the difference of
kinetic and macroscopic scales. 

It is evident that numerical simulations allow to 
study more involved physical models (like full MHD), which cannot be solved 
analytically. On the other hand, the above cited analytical studies,
including the present, have the advantage that they provide exact solutions, 
and they allow us to analyse physical and mathematical effects in great detail.
In particular, analytical investigations are indispensable for studying how the 
distribution of the magnetic null points/stagnation points determines both the 
topology of the streamlines/magnetic field lines and the geometrical shape of the 
heliopause.

In this paper, we provide exact mathematical definitions of astrospheres and astropauses. We 
discuss possible shapes with respect to the spatial arrangement of the stagnation and/or null 
points and emphasize the role of the directions of flow and field of both the ISM and the star.
Furthermore, we discuss the role of magnetic shear flows as possible trigger for particle 
acceleration (e.g., ACRs) in the heliotail.

\section{Geometrical shapes and topological properties of astropauses}

The geometrical shapes of astropauses depend on the strengths and directions of the two involved 
flows (stellar wind and ISM flow) and their electromagnetic fields. The fact that most of the ISM 
flow cannot penetrate the stellar wind region but is forced to flow around defines the 
boundary called astropause. In the mathematical sense, the streamlines in the local ISM
and in the stellar wind region are topologically disjoint. Still, different possibilities
for the definition of the physical astropause exist. One way to define the location of this boundary might be
to use the stagnation point of the flow as criterion. This stagnation point is the intersection
of the stagnation line and the astropause (separatrix) surface. The stagnation line is this streamline along which 
all fluid elements heading towards the astropause and coming from opposite directions, i.e., on the one hand
from the ISM and on the other hand from the star, are 
decelerated down to zero velocity. However, such a definition is only useful in a single-fluid (i.e.
classical HD or MHD) theory, while in a more general multi-fluid model different pauses (i.e. one for 
each component) exist. Here a definition via a 
stagnation point of the flow is not unique. A precise way is thus to use the null point of
the magnetic field, and to transfer the concept of the stagnation point of the flow to the magnetic
field configuration. Such a definition via the null point of the magnetic field makes only sense 
as long as the star itself has a magnetic
field, strong enough to be dynamically important. Here, we focus on stars 
with magnetic fields, for which the astropause can be uniquely defined as the 
outermost magnetic separatrix. The most well-known magetic star is our Sun
with its boundary, the heliopause.

\subsection{Definition of a separatrix}

The field lines of a vector field are typically represented by the trajectories of
a corresponding system of ordinary differential equations, the so-called phase portrait.
Physically interesting phase portraits are those in which null points exist. Then,
a topological classification of the local vector field can be performed by analysing 
the eigenvalues of the corresponding Jacobian matrix of the vector field at the null point 
\citep[see, e.g.,][]{1990ApJ...350..672L, Arnold, 1996PhPl....3..759P}. The configuration of 
the vector field in the vicinity of the magnetic null point depends on the dimension of the 
considered problem.
For instance, in 3D, the magnetic null point is the intersection of the 1D stable or unstable
submanifold termed the spine, with a 2D unstable or stable manifold termed the fan.
This 2D manifold is termed the separatrix (or pause).
In cartesian 2D, the magnetic null point is the intersection of two separatrix field lines of
which one is called stable and the other one is called unstable. 
This null point in 2D is of X-point type. Trajectories, i.e., streamlines or field lines, of the 
vector field are called stable, when the streamline or field line points towards the null 
point, while they are called unstable when they point away from the null point.
Such configurations can only be obtained if the eigenvalues of the Jacobian matrix of the
 vector field at the null point are of so-called hyperbolic or saddle point
type. More specifically, in 2D the eigenvalues are real and have opposite signs, while
in 3D the situation is more complex. Here, the real parts of the eigenvalues should not 
vanish and always two of their real parts have the identical sign. 
Basically, this separatrix concept is valid also concerning the flow field in HD or MHD.

The definition of a separatrix was so far only restricted to a scenario with a single, 
isolated null point. In real astrophysical scenarios, multiple null points may exist, 
depending on the complexity of the stellar wind and its magnetic field. Therefore, multiple and 
highly complex, maybe even nested separatrices may occur.
For instance, \citet[e.g.,][]{2013ApJ...774L...8S} find that the heliopause can be 
considered as a region consisting of bundles of separatrices and magnetic islands, resulting 
from magnetic reconnection processes, which form a porous, multi-layered structure.  
Examples of multiple and nested separatrices originating from X-type null points
are shown and discussed in the following.

\begin{figure*}[t]
\vspace*{2mm}
\centerline{%
\includegraphics[width=0.3\textwidth]{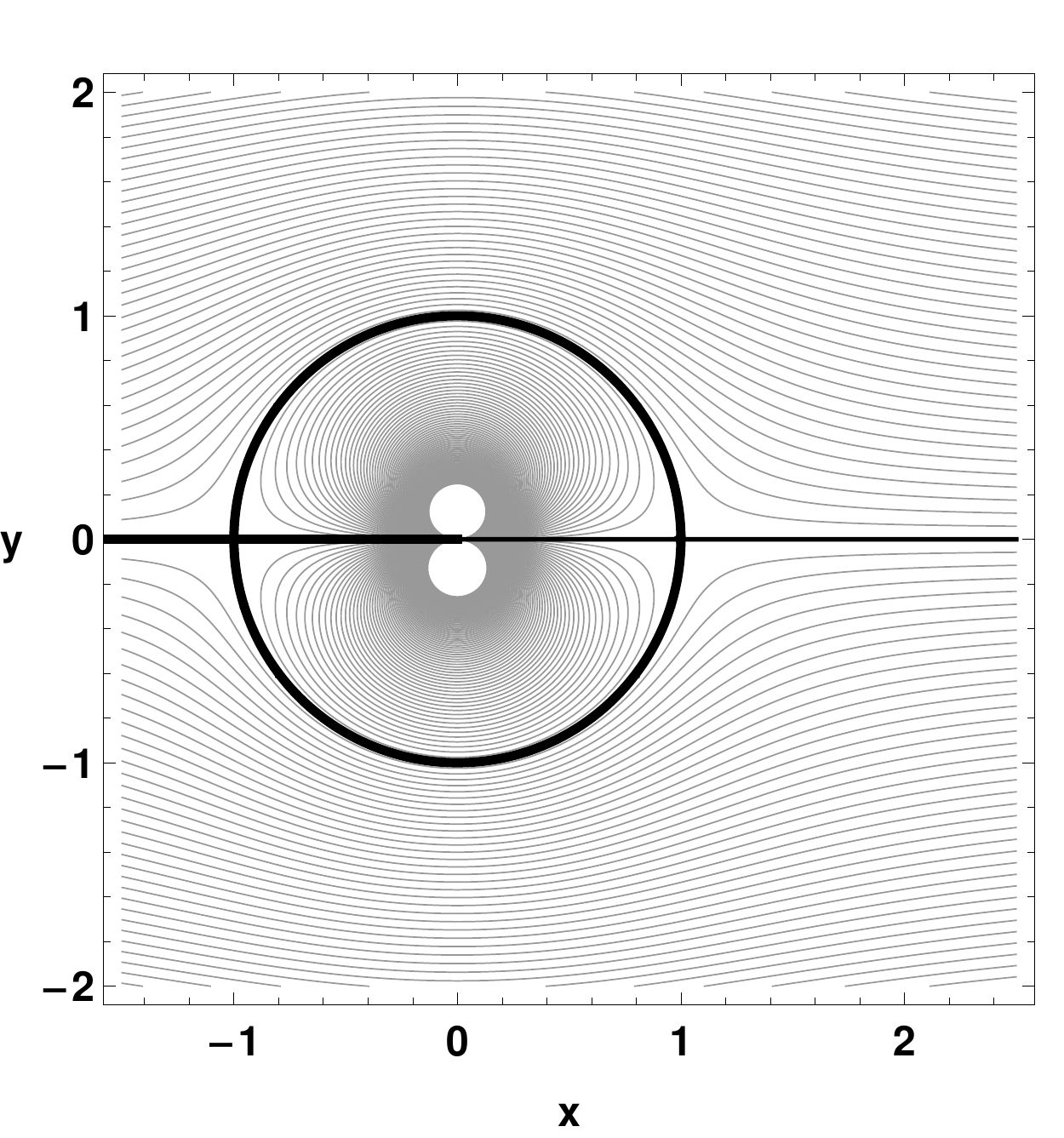}%
\includegraphics[width=0.3\textwidth]{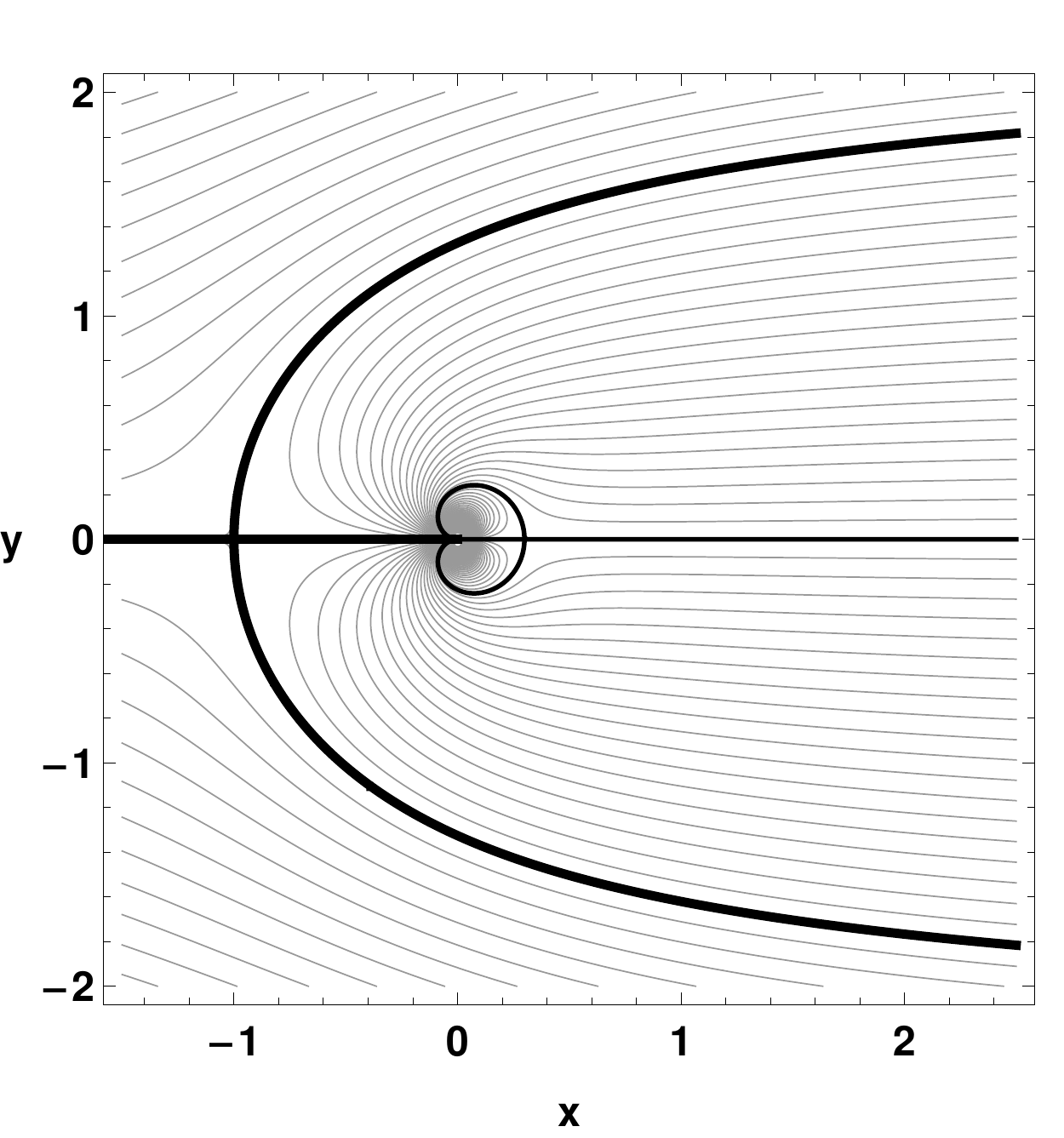}%
}%
\centerline{%
\includegraphics[width=0.3\textwidth]{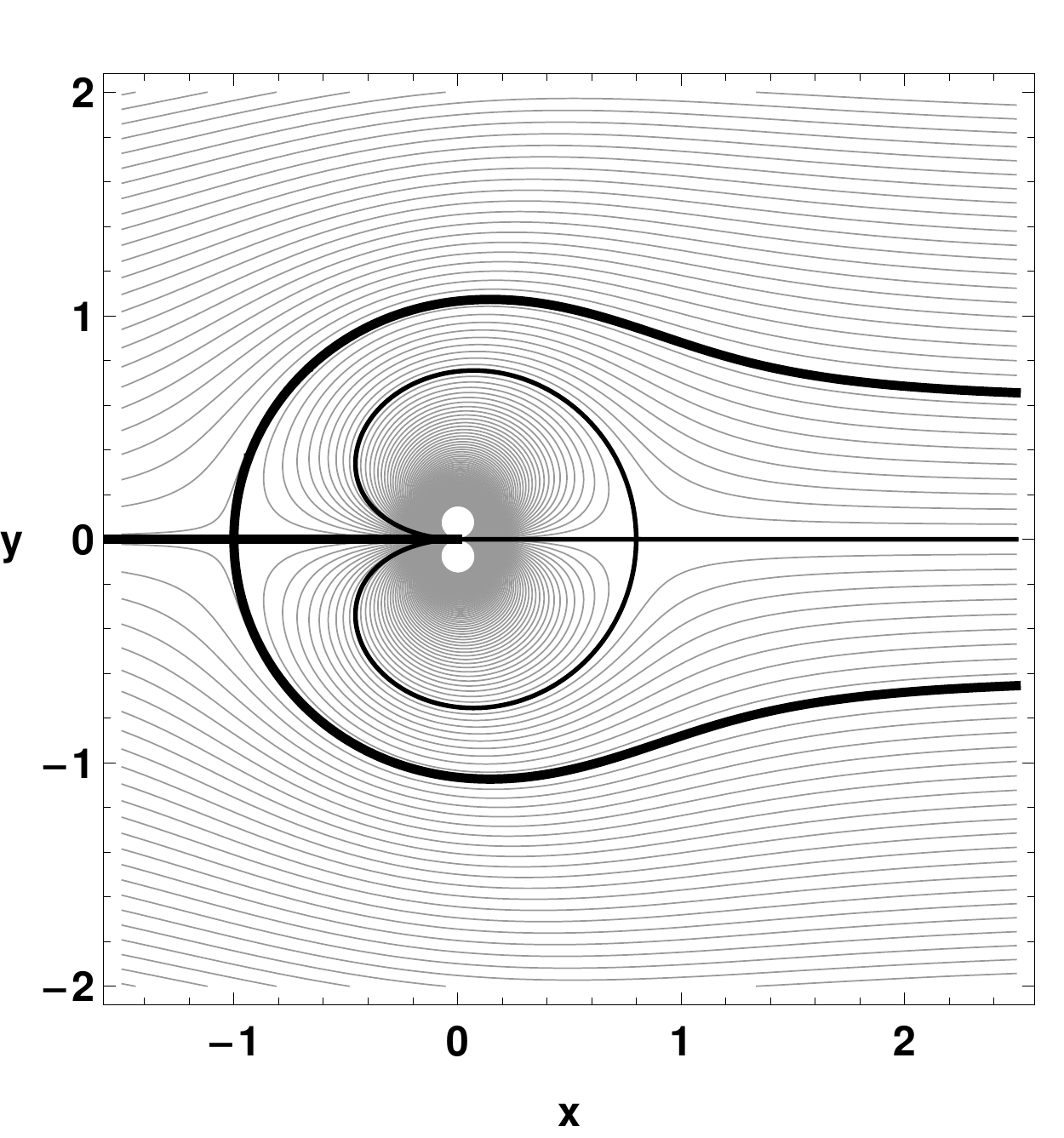}%
\includegraphics[width=0.3\textwidth]{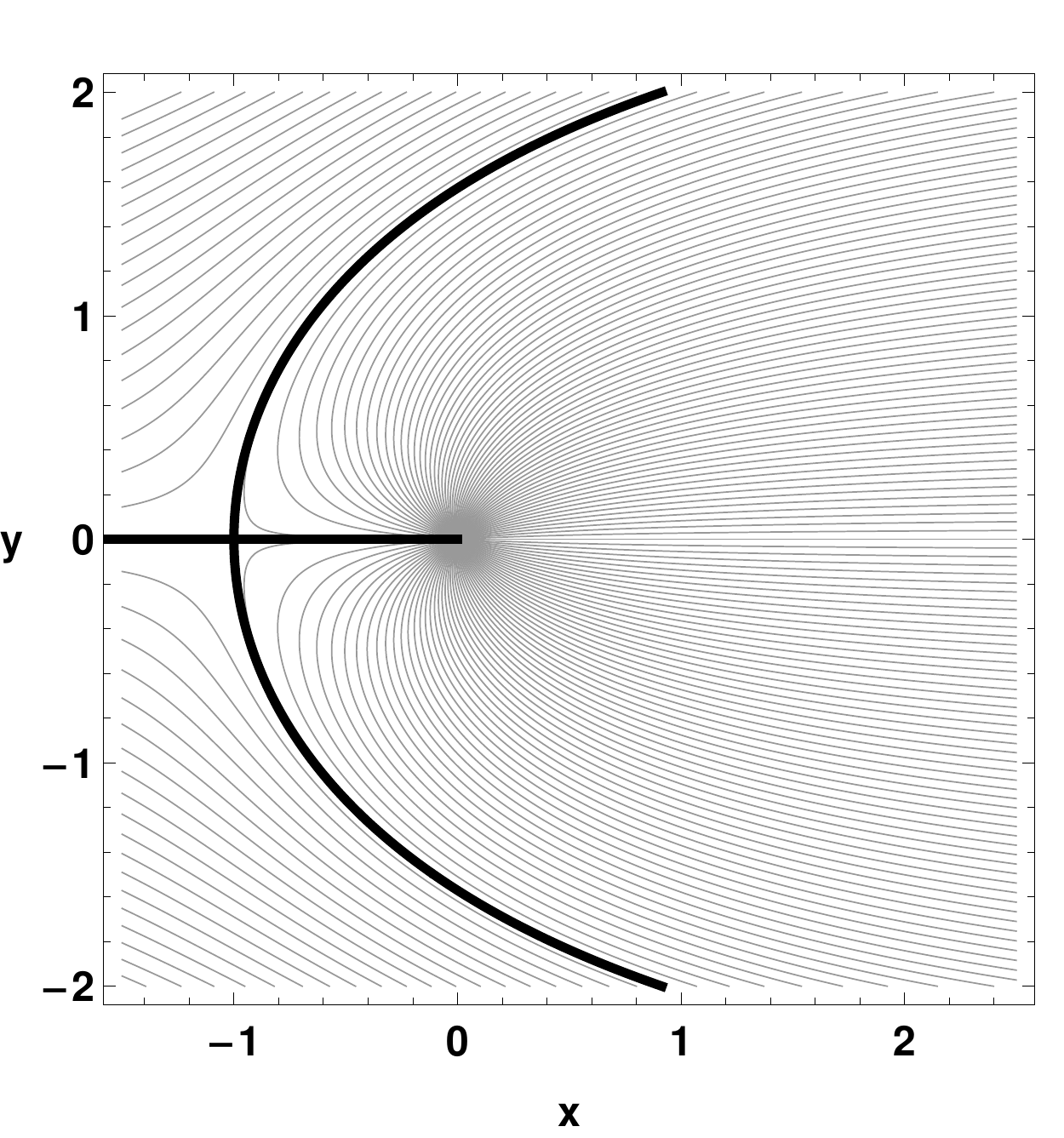}%
}%
\centerline{%
\includegraphics[width=0.3\textwidth]{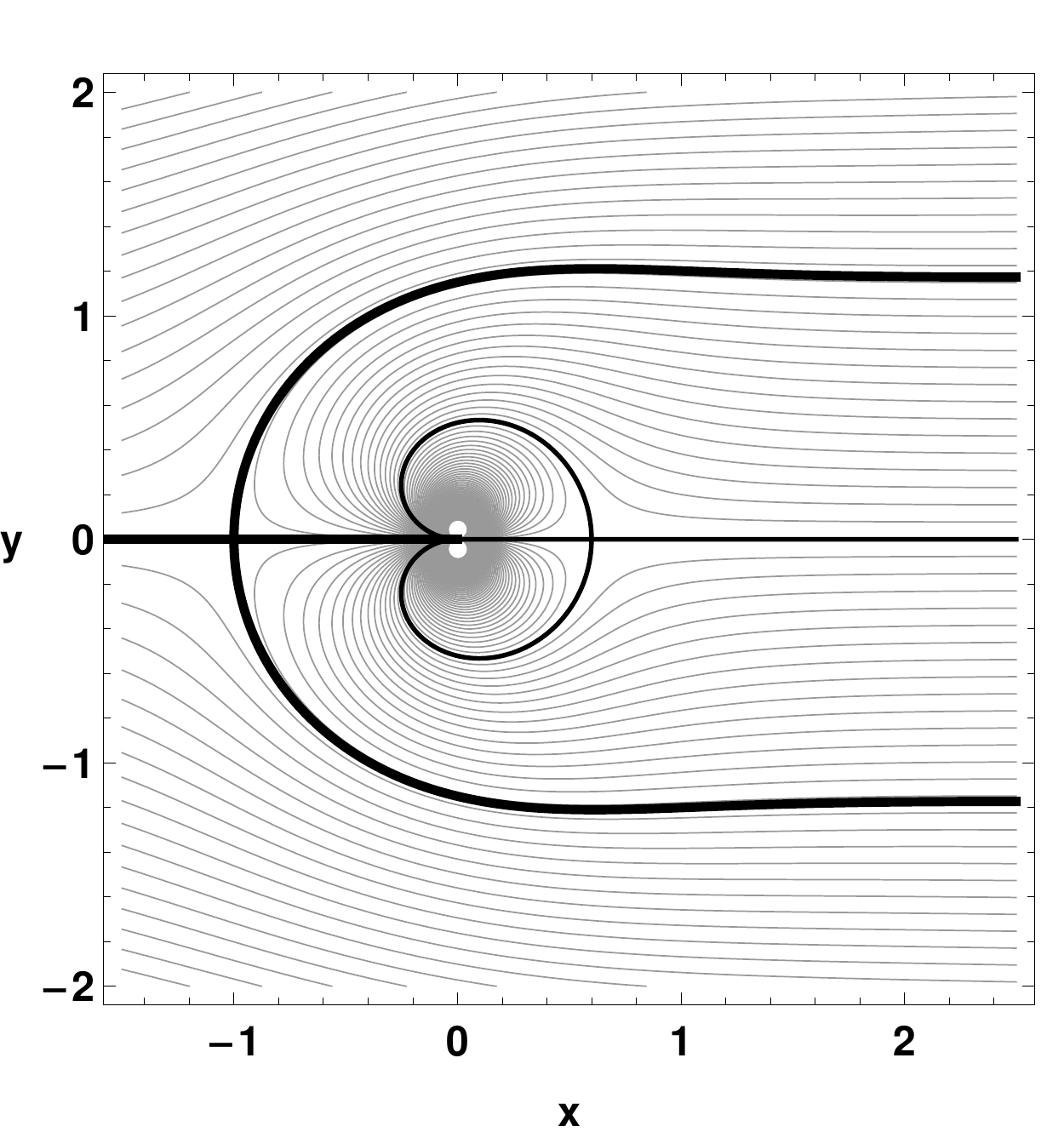}%
\includegraphics[width=0.3\textwidth]{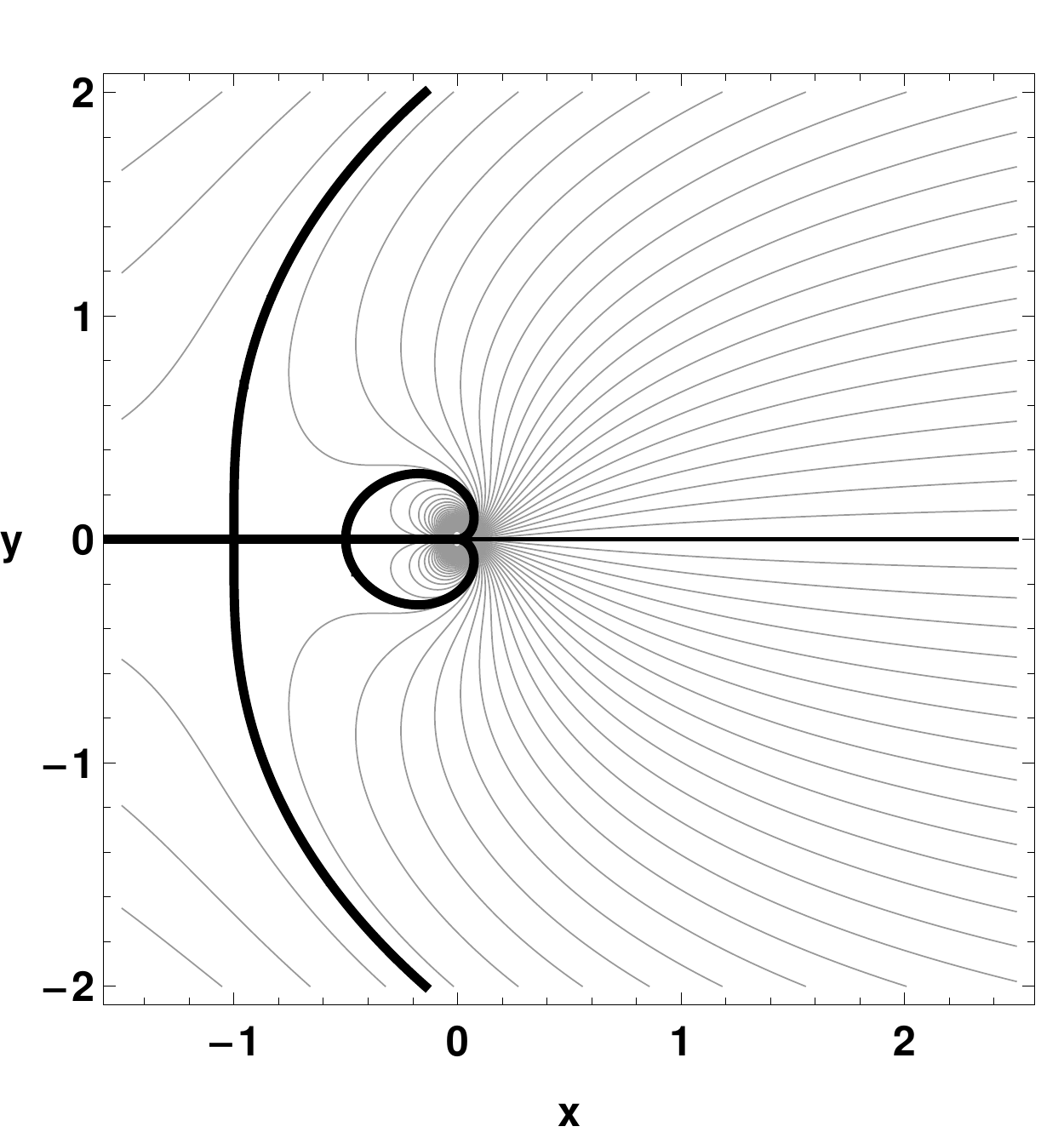}%
}%
\caption{Shape of the separatrices in dimensionless units 
for the case of two null points lying on
the $x$-axis. Shown are the field lines (i.e., the projection of
the contour lines of $A$ into the $x-y$-plane).
One null point is fixed at $u_{1} = x = -1$, the other one, $u_{2}$,
is at $x=1$ (symmetric, left top), $x=0.8$ (left middle), $x=0.6$ (left bottom),
$x=0.3$ (right top), $x=0$ (Parker scenario, right middle), $x=-0.5$ (right
bottom). In the Parker scenario, the second null point disappears (coincides
with the pole) and therefore also the dipole moment. The separatrices defined
by $u_{1}$ are plotted with strongest, those defined by $u_{2}$ with medium,
and field lines are shown with normal line width.}
\label{series}
\end{figure*}

\subsection{Determination of the global shape of astrospheres: distribution of null points
for pure potential fields}

To compute non-linear MHD flows with separatrices as we will do in Sect.\,\ref{flows}, 
geometrical patterns are needed for
the structure of the corresponding flows and fields. To obtain such patterns, we start from 
the simplest possible fields, the potential fields, and investigate how
separatrices form and how they are shaped by different spatial distributions of null points.
The advantage of using potential fields is given by the fact that these fields obey a 
superposition principle, i.e., they are linear, and every null point is automatically
an X-point of the magnetic field. In addition, potential fields have no free magnetic 
energy (i.e., they are stable) and can easily be mapped
to non-linear fields by either generalized contact transformations or algebraic transformations. 
The concept of contact transformations was presented, e.g., by \citet{1992PhFlB...4.1689G} and 
later on used and refined by \citet{2010AnGeo..28.1523N, 2012AnGeo..30..545N, 
2006A&A...454..797N, 2013A&A...556A..61N} to describe the MHD fields of different space plasma 
environments, while the algebraic transformations were introduced by 
\citet{2000JMP....41.2043B, 2000PhRvE..62.8616B, 2001PhLA..291..256B, 2002PhRvE..66e6410B}.

As was shown by \citet{1993A&A...277..249F} in the case of pure HD and by \citet{2006A&A...454..797N}
in the case of MHD, the number and spatial distribution of the hyperbolic null points determine the 
topological scaffold of an astrosphere. This, in combination with the assumption of 
a homogeneous background field as asymptotic boundary condition, allows us to describe the 
general shape of its field and streamlines.
In the current paper, we aim at giving a qualitative overview of the different scenarios. 
For the detailed theoretical description and treatment we refer to \citet{2006A&A...454..797N}.
 
If several hyperbolic null points exist, which of the separatrices defines then the real pause? 
What can be said about multiple null points is that they all have to be non-degenerate, because
double or higher order null points are topologically unstable \citep[e.g.,][]{1996PhPl....3..781H}.

Having discussed the general topological aspects, we now focus on the geometrical ones, 
considering a simplified 2D scenario, in analogy to typical flows in aero and fluid dynamics.  
While in the vicinity of the star the fields
are full 2D to account for a variety of multipolar field structures, asymptotically,
i.e., far away from the star in downwind direction, the field converges to a tail-like (1D) 
structure.
Let us start with the case of two null points in the frame of a 2D cartesian potential field. 
We assume that the $z$-direction is the invariant direction, which means that for all 
parameters $\partial/\partial z = 0$. Further, $x$ points to the downwind (i.e. tail) direction,
and $y$ in perpendicular direction. The global magnetic field, $\vec B = \vec\nabla A(x,y)
\times \vec e_{z}$, with the unit vector $\vec e_{z}$ in $z$-direction, can be described via
complex analysis. 
As $\vec B$ should be a potential field, it follows that $\Delta A = 0$. To solve this Laplace
equation, we define a stream or, here, the magnetic flux function $A$ by $A= \Im({\cal A})$,
where ${\cal A}$ is the complex stream or magnetic flux function. This complex flux function
is obtained from a Laurent series of the form \citep[see, e.g.,][]{2006A&A...454..797N} 
\begin{equation}
{\cal A} = B_{S\infty} u + C_{0} \ln u + \frac{C_{1}}{u} + \textrm{terms of higher order}\, . 
\end{equation}
In the following, we will neglect the higher order terms, so that
the complex magnetic flux function consists purely of a monopole\footnote{The 
monopole is introduced as a mathematical tool, used to generate radial streamlines 
(i.e. the stellar wind) and radial (i.e. open) field lines. Without it, the streamlines 
and magnetic field lines would otherwise always be closed.} and dipole.
Both multipoles are located in the origin.
Here, $u= x + i y$ is the complex coordinate, $B_{S\infty}$ is the asymptotical boundary 
condition $\lim B = B_{S\infty}$ for $|u|\rightarrow \infty$, i.e., the background field,
and $C_{0}$ and $C_{1}$ are the monopole and dipole moments, respectively.
For the case of two null points, these moments have the following form
\begin{equation}
C_{0} = -B_{S\infty} (u_{1} + u_{2}) \qquad \textrm{and} \qquad C_{1} = -B_{S\infty} u_{1} u_{2}
\end{equation}
where $u_{1}$ and $u_{2}$ are the complex coordinates of the two null points.

\begin{figure}[t]
\vspace*{2mm}
\begin{center}
\includegraphics[width=8.3cm]{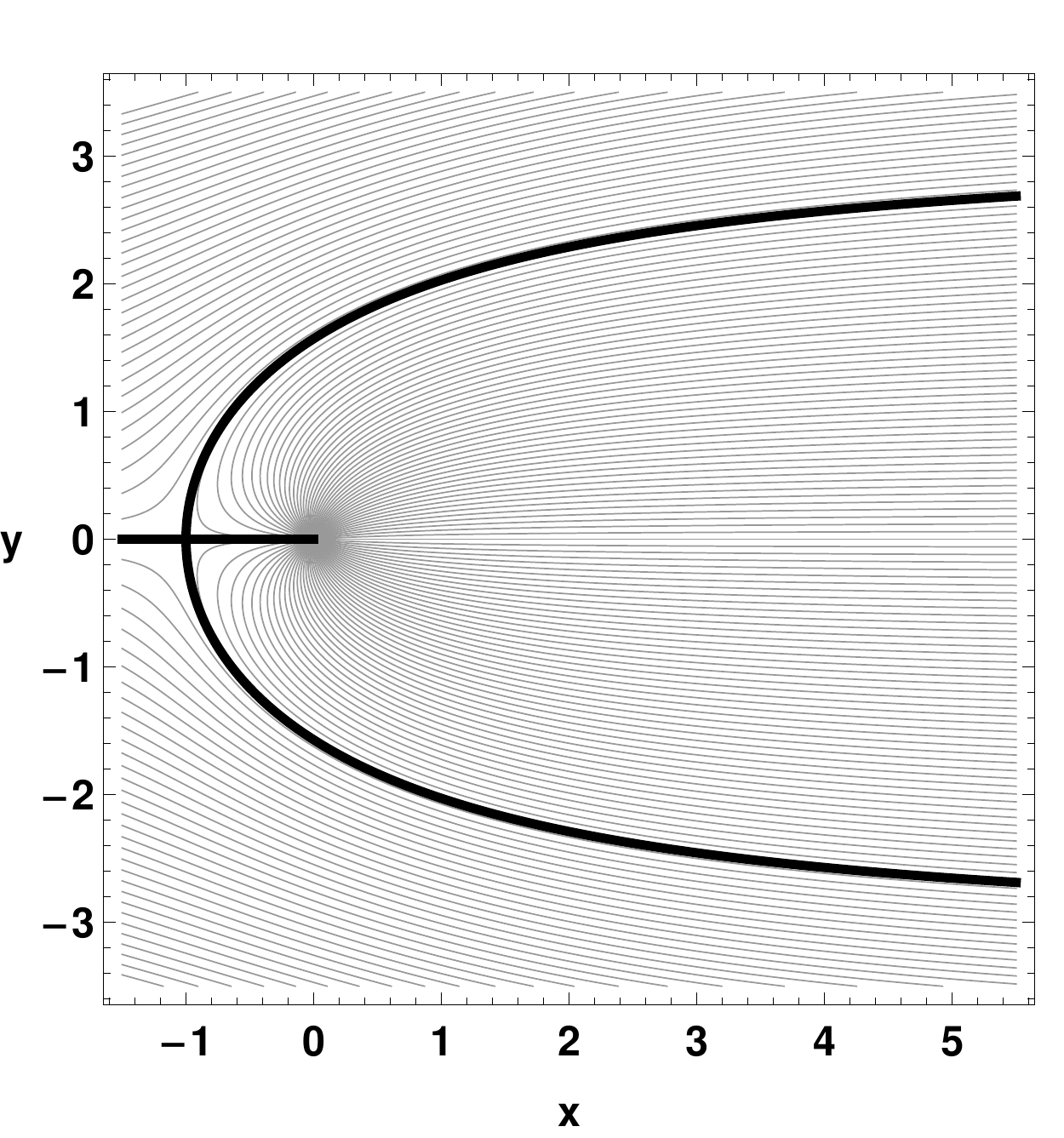}
\includegraphics[width=8.3cm]{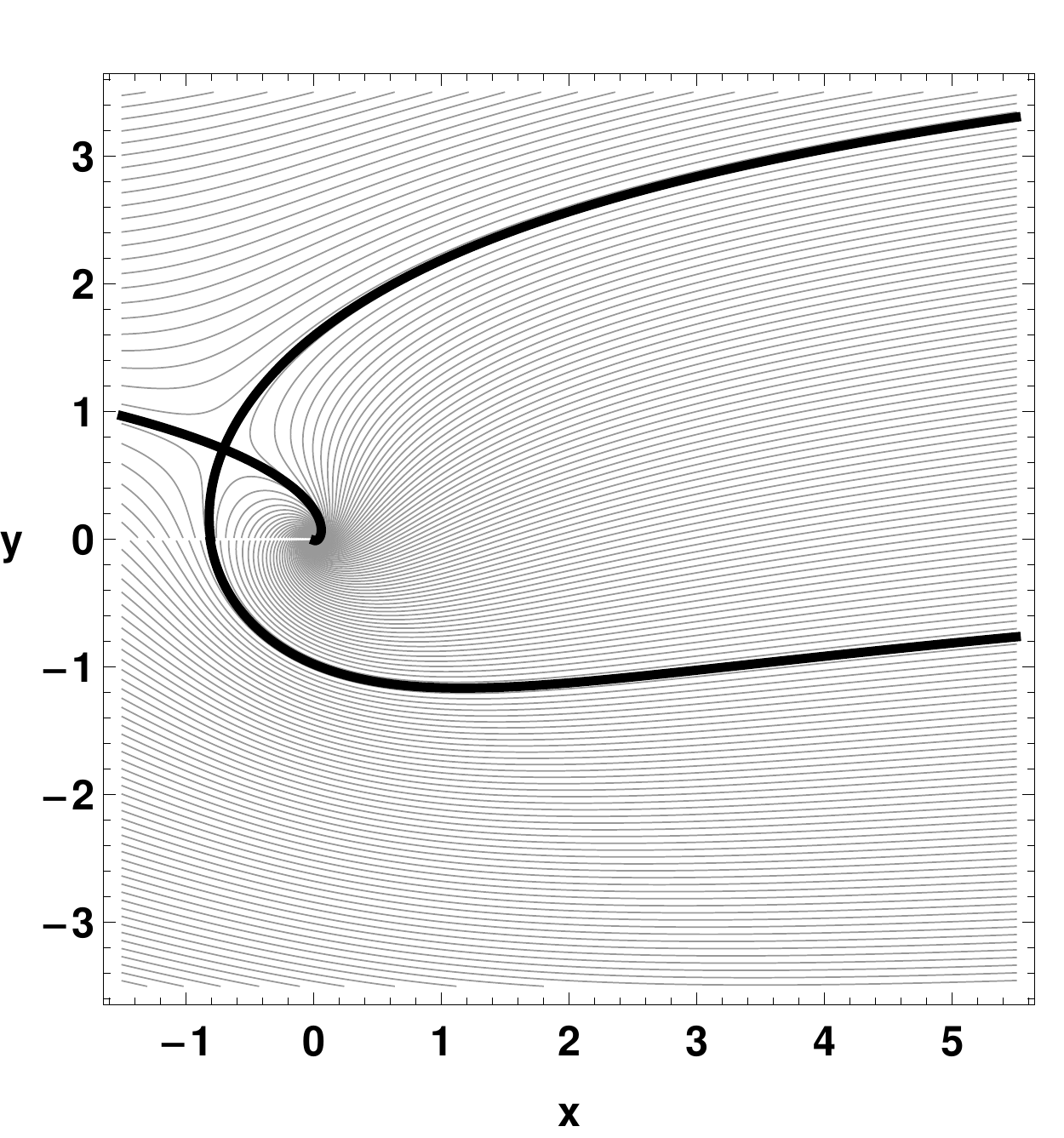}
\end{center}
\caption{As Fig.\,\ref{series}, but comparing separatrix shapes with only one null point. Top: the symmetric Parker case
($u_{1} = x = -1$), and bottom: the asymmetric case where $u_{1}$ is rotated off the $x$-axis by
$-\pi/4$, i.e., $u_{1} = -1 (\cos(-\pi/4)+ i \sin(-\pi/4))$. The apparent
non-connectivity of some field lines at $y=0$ and $x < 0$ is not real but caused by the flip between
Riemann surfaces.}
\label{axisym}
\end{figure}

\begin{figure}[t]
\vspace*{2mm}
\begin{center}
\includegraphics[width=8.3cm]{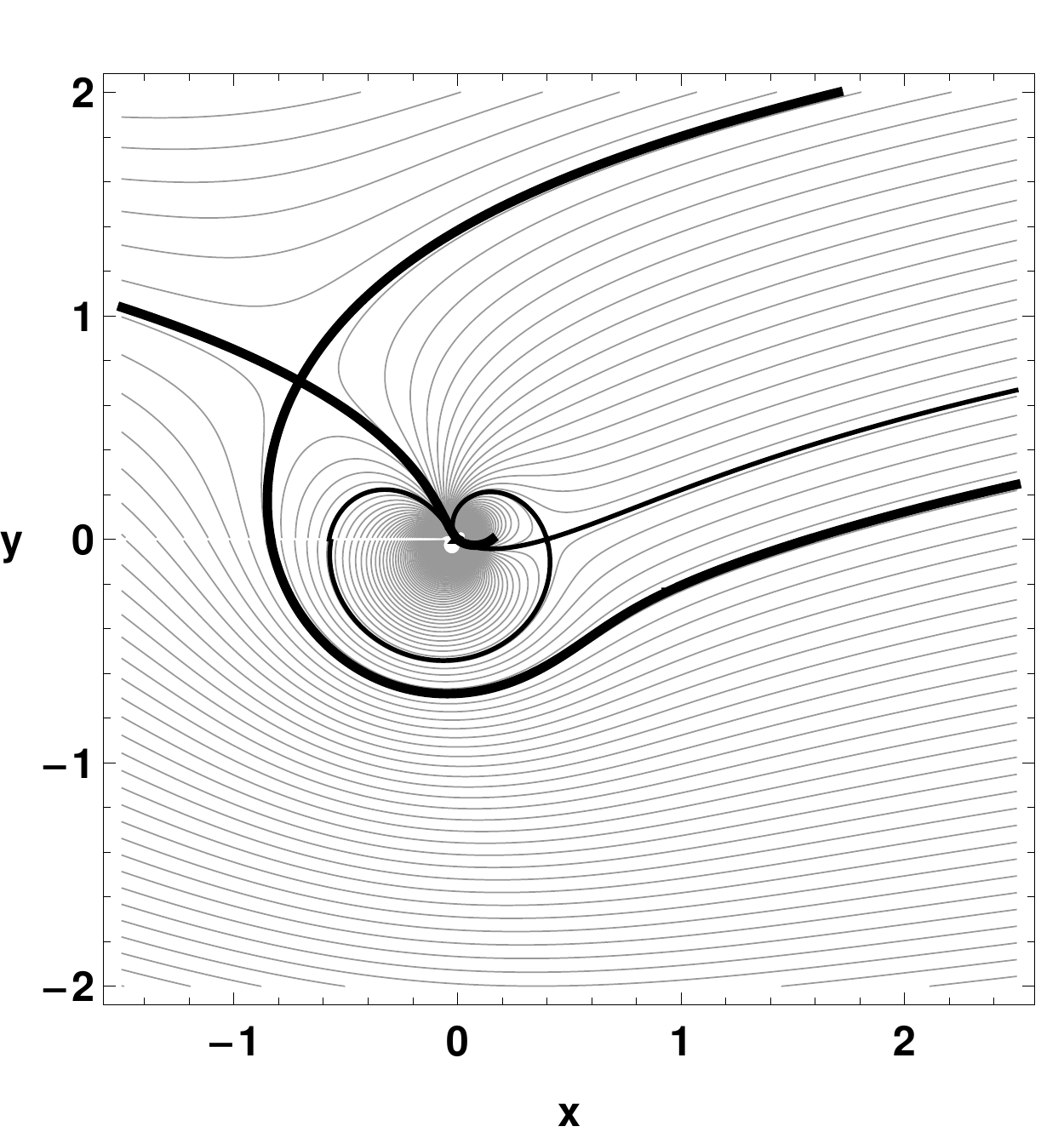}
\end{center}
\caption{As Fig.\,\ref{series}, but for an asymmetric case where the front null point
is rotated off the $x$-axis by $-\pi/4$, i.e., $u_{1} = -1 (\cos(-\pi/4)+
i \sin(-\pi/4))$, and the other one is on the $x$-axis at $u_{2} = x=0.4$.
The apparent non-connectivity of some field lines at $y=0$ and $x < 0$ is not 
real but caused by the flip between Riemann surfaces.}
\label{twist}
\end{figure}

The simplest scenario is the one with two symmetric null points ($u_{1}= -u_{2}$).
In this case, only the dipole moment exists, and we can interprete this with a star with a 
dipole magnetic field embedded
in a homogeneous magnetic background field. Such a scenario is an analogy to the 
classical hydrodynamical example of a cylindrical obstacle in a homogeneous flow. 
If we assume the star is located at the origin of a cartesian coordinate system and the flow 
is parallel to the $x$-axis and streams in positive $x$-direction, the separatrices form 
a circle in the $(x,y)$-plane, where the stagnation lines lie on the 
$x$-axis and intersect the circular separatrix from both sides at the two null points and pass through 
the pole. This is shown in the upper left panel of Fig.\,\ref{series}. The radius $R$ of the circular 
separatrix, and hence the location of the two symmetric null points, depends on the strength of  
the background magnetic field ($B_{S\infty}$) and the dipole field via $R^{2} = B_0 R_0^2 / B_{S\infty}$.
The term $B_0 R_0^2$ is hereby the dipole moment. 
Such a symmetric scenario is not very realistic, because it 
would imply a completely closed separatrix. Hence, no plasma can escape via the stellar wind.

To enable at least a half-open astrosphere, the second null point (the one in the downwind
region) must be located closer to the pole. The shape of the resulting separatrices for different
pole distances are depicted in the series of plots in Fig.\,\ref{series}. 
When moving the second null point towards the pole, one
can notice several effects. First, the separatrix resulting from the first null point \lq opens\rq \ in 
the downwind region. Second, an inner, closed separatrix is formed that passes through the second null 
point and through the pole. This separatrix encloses now the dipolar field region, which became smaller, 
and the dipole field weaker, if the same background field is present. This inner separatrix meets the 
stagnation line from the first null point at the pole. A measurable bundle of field lines can leave the inner 
dipole region upstream, i.e., between the inner separatrix and the stagnation line. These field lines 
origin from the monopole part of the field. They are deflected at the outer separatrix so that they bend 
around the inner separatrix and extend as open field lines into the astrotail, forming the inner 
astrosheath field lines. The closer the second null point is located to the pole, the more opens the 
tail region.

As soon as the second null point \lq reaches\rq \ the pole, it vanishes. Hence, only one real null point 
remains and the field has only a monopole moment. This scene is shown in the 
middle right panel of Fig.\,\ref{series} and is similar to the Parker scenario 
\citep{1961ApJ...134...20P} for subsonic flows. If the second null point is located on the same
side as the first one, i.e., in upwind direction (lower right panel of Fig.\,\ref{series}), 
the second separatrix is also located in upwind direction and both null points are physically 
connected by the stagnation line. In addition, the monopole moment increases so that the
astrotail becomes even wider.  

Restricting for the moment to a single null point, it is, of course, not necessary that this null 
point is located on the $x$-axis. For instance, with respect to the heliopause, measurements from 
{\it Voyager 1} indicate an asymmetry \citep{2013Sci...341..147B}. A natural way to displace 
the null point is provided by the solar (or stellar, in general) rotation, 
which results in a winding-up of the field lines and hence
to a spiral structure of the magnetic field. While the monopole moment in the symmetric examples
is a pure real number, it now becomes complex. Hence, an azimuthal component of the outflow or field
occurs. The result is a displacement of the null point off the $x$-axis. This is demonstrated in 
Fig.\,\ref{axisym}, where we plot the asymmetric configuration\footnote{Considering that 
{\it Voyager 1} might have passed the stagnation region at a distance of 123\,AU 
\citep{2013Sci...341..144K}, meaning that within our scenario $u_{1}$ would be at roughly 
123\,AU, our configuration shown in Fig.\,\ref{axisym} (bottom) and Fig.\,\ref{twist} 
might be approximately scaled with 1:87\,AU.} in comparison to 
the symmetric one. An even more complex situation is achieved when a second null point exists 
in the asymmetric scene. Such an example is shown in Fig.\,\ref{twist}. 
   
Having multiple, nested separatrices, the real astropause can be uniquely defined by the 
outermost magnetic 
separatrix between the magnetized interstellar medium and the magnetized stellar wind.

\section{Self-consistent non-linear MHD flows}\label{flows}

The pattern of the potential fields, as we calculated in the previous section, serve
now as static MHD equilibria (MHS). These are then mapped with the non-canonical transformation
method to self-consistent steady-state MHD flows. Thereby we make use of estimated or
observed physical quantities, such as density, magnetic field strength, etc., within
and outside the heliosphere. These quantities serve as asymptotical boundary conditions,
based on which some of the coefficients of the mapping can be fixed. 
   
Observations from {\it Voyager 1} suggest that the plasma flow in the vicinity of the 
heliopause and in the heliotail region is approximately
parallel to the magnetic field \citep{2013Sci...341..147B, 2013ApJ...776...79F}. 
In addition, the plasma within a stagnation region is incompressible. 
This can be understood in terms of the steady-state mass continuity equation 
\begin{eqnarray} 
\vec\nabla\cdot\left(\rho\,\vec v\right)=0 
\quad\Leftrightarrow 
\quad \vec v\cdot\vec\nabla\rho + \rho \vec\nabla\cdot\vec v=0 \, . 
\end{eqnarray} 
When approaching the stagnation point, i.e. $\vec v\rightarrow \vec 0$, the 
term $\vec v\cdot\vec\nabla\rho$ in the second equation vanishes, implying that
$\rho\vec\nabla\cdot\vec v$, and in particular, as the density reaches a maximum, 
$\vec\nabla\cdot\vec v$ has to vanish as well.\footnote{This remains valid, even if 
$\vec\nabla\rho$ happens to become extremely large across the heliopause boundary layer.} 
The plasma flow on streamlines, which originate in such stagnation point regions, 
transports the property of incompressibility further into the tail region.
Hence, it is reasonable to investigate the heliotail and heliopause region using
field-aligned, incompressible flows, and the basic ideal MHD equations are given by
\begin{eqnarray}
  \vec\nabla\cdot\left( \rho\vec{\rm v} \right) &=& 0\, ,\label{mce}\\
   \rho\left( \vec{\rm v}\cdot\vec\nabla\right)\vec{\rm v} &=& \vec
j\times\vec B - \vec\nabla P\, \label{ee},\\
\vec\nabla\times\left(\vec{\rm v}\times\vec B\right) &=&\vec 0\, ,\label{ie}\\
     \vec\nabla\times\vec B &=& \mu_{0}\vec j\, ,\label{al}\\
      \vec\nabla\cdot\vec B &=& 0\, ,\label{sc}\\
       \vec\nabla\cdot\vec{\rm v} &=& 0\, ,\label{ice}\\
       \vec{\rm v} &=& \pm|M_{A}| \vec{\rm v_{A}}  \\ 
      \vec{\rm v_{A}}& := &\frac{\vec B}{\sqrt{\mu_{0}\rho}}\, , \label{letzte}
        \end{eqnarray}
where $\rho$ is the mass density, $\vec{\rm v}$ is the plasma velocity, $\vec B$ is the magnetic flux density,
$\vec j$ is the current density, $P$ is the plasma pressure, $M_{A}$ is the Alfv\'{e}n Mach number,
$\vec{\rm v_{A}}$ is the Alfv\'{e}n velocity, and $\mu_{0}$ is the magnetic permeability of the vacuum. 

Given solutions for $p_{S}$ and $B_{S}$ of the MHS equations 
\begin{equation}
\vec\nabla p_{S}  =  \vec j_{S}\times\vec B_{S}\, , 
\end{equation}
and additional solutions for $M_{A}$ and $\rho$ of the systems
\begin{eqnarray}
\vec B_{S}\cdot\vec\nabla\rho & = & 0\, , \\
\vec B_{S}\cdot\vec\nabla M_{A} & = & 0\, ,
\end{eqnarray}
are the parameters needed to perform the transformation, i.e., to compute the 
general solution \citep[see, e.g.,][]{2012AnGeo..30..545N} of the system Eq.\,(\ref{mce})-(\ref{letzte})
\begin{eqnarray}
\vec B &= &\frac{\vec B_{S}}{\sqrt{1-M_{A}^2}}\, , \\ 
p & = & p_{S} - \frac{1}{2\mu_{0}}\frac{M_{A}^2\left|\vec B_{S}\right|^2}{1-M_{A}^2}\, , \label{magpressuretrafo}\\
\sqrt{\rho} \vec{\rm v} &= & \frac{1}{\sqrt{\mu_{0}}}\,
\frac{M_{A}\vec B_{S}}{\sqrt{1-M_{A}^2}}\, ,\\
\vec j & = &  \frac{M_{A}}{\mu_{0}}\frac{\vec\nabla M_{A}\times\vec B_{S}}{\left(1-M_{A}^2
\right)^{\frac{3}{2}}} +\frac{\vec j_{S}}{\left(1-M_{A}^2\right)^{\frac{1}{2}}}\, . \label{strom}
\label{streaming2mhs1}
\end{eqnarray}
Properties of these solutions are that the plasma density $\rho$, the Alfv\'{e}n Mach number $M_{A}$, and
the Bernoulli-pressure $\varPi = P+\frac{1}{2}\rho \vec {\rm v}^2$ are constant on field lines. However,
these parameters can vary perpendicular to the field lines, which means that for example strong shear
flows, implying a strong gradient of the Alfv\'{e}n Mach number, produce strong current
densities (curent sheets). This is obvious from Eq.\,(\ref{strom}), even if we start from a potential
field, which implies $\vec j_{S} = \vec 0$. The occurrence of current sheets, especially around multiple
separatrices, is known, e.g., in solar flare physics, as current fragmentation 
\citep[e.g.,][]{2008CEAB...32...35K, 2008SoPh..247..335K, 2010AdSpR..45...10B},
or, in steady-state as fragmented currents \citep[e.g.,][]{2013A&A...556A..61N}.
It should be emphasized that a similar transformation can also be performed
using super-Alfv\'{e}nic flows.
 
In the following, we concentrate on the tail region, taking the symmetric Parker-like
tail (top panel of Fig.\,\ref{axisym}).
The mapping ansatz we use was described in detail in \citet{2006A&A...454..797N}.
The Alfv\'{e}n Mach number is defined via $M_{A}^{2} = 1-1/(\alpha'(A))^{2}$, where the prime 
denotes the derivation with respect to $A$, and $\alpha$ is the mapped $z$-component of the 
vector potential of $A$. This means that if $\vec B_{S} = \vec\nabla A \times \vec e_{z}$
and $\alpha = \alpha(A)$, then $\vec B = \vec\nabla\alpha(A) \times \vec e_{z} = \alpha'(A)
\vec\nabla A \times \vec e_{z}$. Hence, $\alpha'(A)$ is the amplification factor for the
magnetic field strength, and it results to 
\begin{eqnarray}
\alpha'(A) & = & \frac{1}{2}\left( \frac{1}{\sqrt{1-M_{A\infty}^2}} - \frac{1}{\sqrt{1-M_{A,i}^2}} \right)\\
&  & \cdot  \left(\tanh{\frac{\frac{A}{\sqrt{1 - M_{A\infty}^2}\, B_{\infty}}
    - y_{1}}{d_{1}}} -\tanh{\frac{\frac{A}{\sqrt{1 - M_{A\infty}^2}\, B_{\infty}}+
y_{1}}{d_{1}}}\right)\, .\nonumber
     \end{eqnarray}
The ansatz for $\alpha'(A)$ is chosen in order to mimic two current sheets located symmetrically
at $\pm y_{1}$
around the heliopause field lines, with oppositely directed currents. 
To compute the Mach number profile and the current sheets, we use the following set of values
\citep[see, e.g.][]{2012ApJ...760..106F, 2013Sci...341..147B}:
For the magnetic field of the interstellar medium $B_{\infty} = 5\,\mu$G, for the 
inner magnetic field strength in the tail $B_{\rm i} = 4\,\mu$G, the particle number densities
of electrons and ions are about equal and $n_{\rm e} \approx n_{\rm i} = 0.1$\,cm$^{-3}$,
the velocity of the ISM plasma relative to the sun is ${\rm v}_{\infty} = 25$\,km\,s$^{-1}$,
and the Mach numbers of the ISM plasma and of the heliotail result to $M_{\rm A\infty} =
0.72$ and $M_{\rm A,i} = 0.52$. These values might not be absolutely true, but they are
used here to provide a rough idea of the global shape of the current sheets.

\begin{figure}[t]
\vspace*{2mm}
\begin{center}
\includegraphics[width=7.5cm]{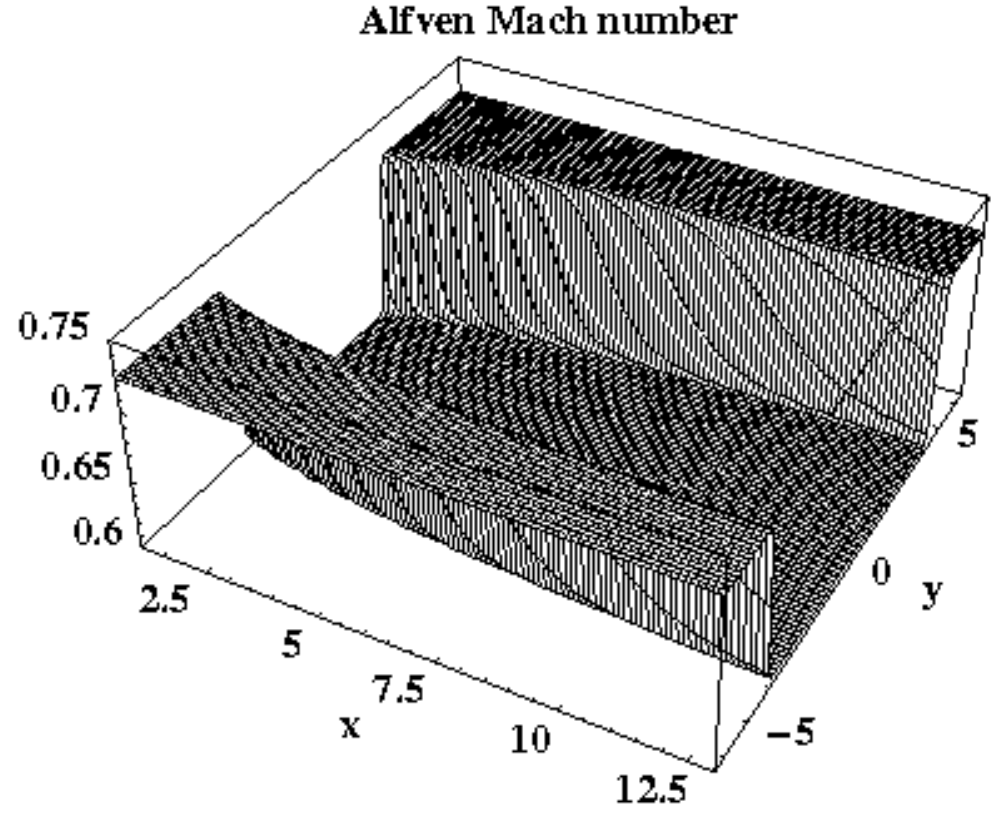}
\includegraphics[width=7.5cm]{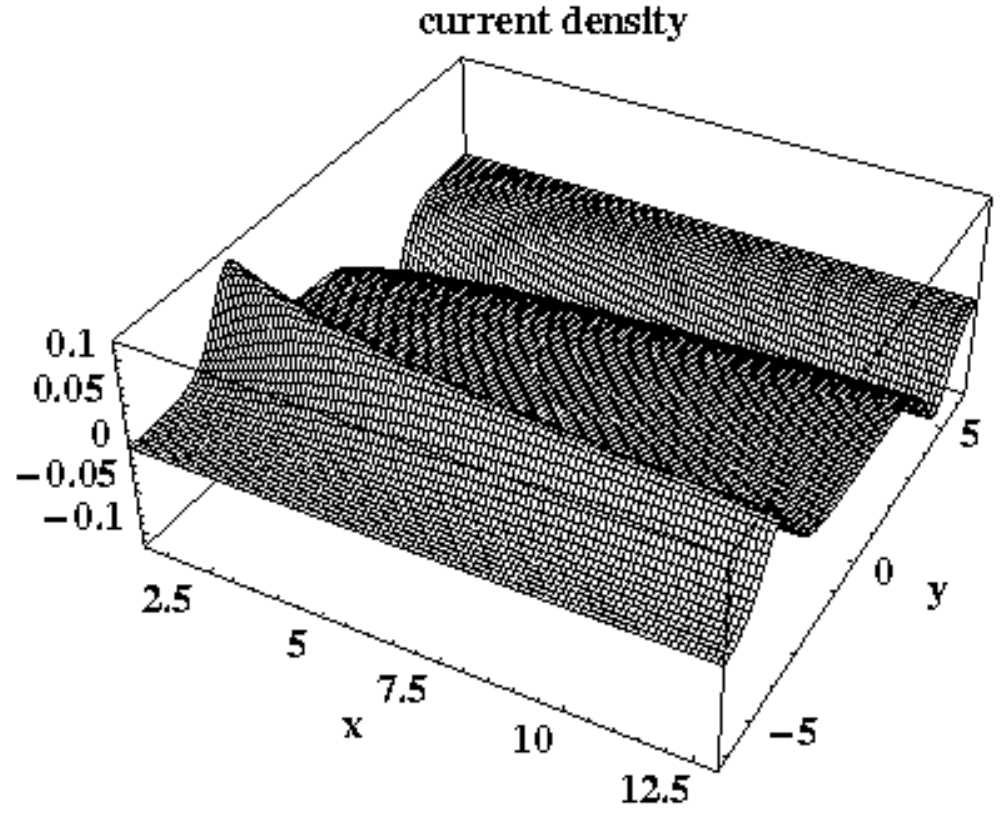}
\includegraphics[width=7.5cm]{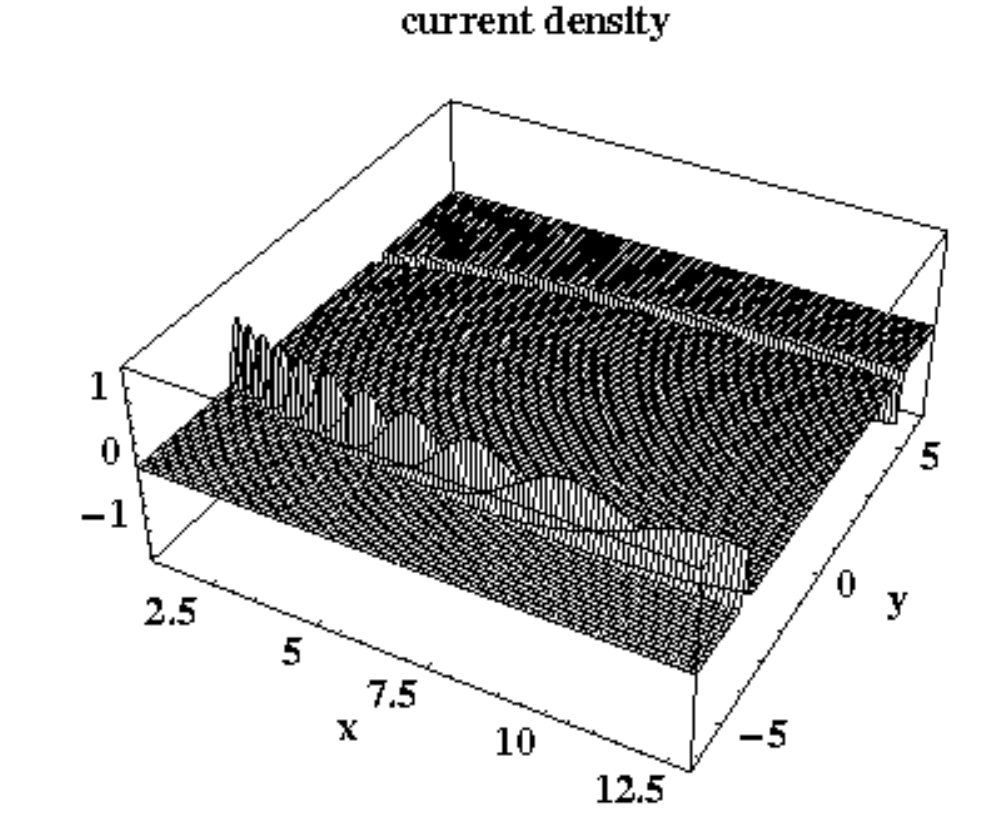}
\end{center}
\caption{Alfv\'{e}n Mach number profile (top, restricted to $M_{\rm A} > 0.6$ 
for displaying purposes) and resulting current sheets for a
sheet width of 100\,AU (middle) and 10\,AU (bottom). The current density of the
poloidal magnetic field is here given in units of $2.65\times 10^{-17}$\,A\,m$^{-2}$.}
\label{currentsheets}
\end{figure}

The Alfv\'{e}n Mach number profile for the chosen parameters is shown in the top panel
of Fig.\,\ref{currentsheets}. It is computed for a thickness of the current sheets in the tail
of $d_{1} = 100$\,AU. This is an unrealistic scenario, and is only shown to highlight the current 
distribution in the tail (middle panel of Fig.\,\ref{currentsheets}). A more realistic
case for the current sheets requires much narrower widths. This is depicted in the lower
panel of Fig.\,\ref{currentsheets} where we use a width of only 10\,AU. Obviously, a reduction
in the width by a factor of ten leads to an increase of the current strength by a factor of
ten. At the boundary between ISM and solar wind flows, Kelvin-Helmholtz-like or current driven 
reconnection instabilities can occur due to shear flows and extremely high current densities, 
respectively. These instability regions are thus ideal locations for 
plasma heating and particle acceleration.

\section{Magnetic shear as trigger for particle acceleration}

The separatrix regions and, in particular, the heliotail regions can serve 
as important particle acceleration locations. This assumption is supported by observations
of both the cosmic ray anisotropy and the broad excess of sub-TeV cosmic rays in the direction
of the heliotail. \citet{2010ApJ...722..188L} propose that this excess originates from 
magnetic reconnection in the magnetotail. 
The heliotail/heliopause region was also considered by \citet{2009ApJ...703....8L} 
as an important region for particle acceleration. In their approach, 
\citet{2009ApJ...703....8L} use a Spitzer-like 
resistivity and propose first-order Fermi acceleration as the dominant 
acceleration process of energetic particles along the magnetotail. 
In contrast, to approach the problematics of particle acceleration we focus on the magnetic
shear generated by magnetic shear flows across the heliopause boundary. 
To reduce the complexity of the problem, we shear here the magnetic field 
only in $z$-direction. This guarantees that the current, and, therefore, also the electric
field, are aligned with the tail direction.
 
We investigate the generation of parallel electric fields and 
consider them as acceleration engines which, 
in solar physics, is typically referred to as Direct Current (DC)
field acceleration \citep[see, e.g.,][]{2002SSRv..101....1A}.
The link to solar physics scenarios is obvious, as the acceleration
process takes place in a diluted plasma environment, i.e., in regions
of low density.

As the magnetic field jumps across the heliopause, the magnetic shear in $z$-direction,
$B_{z}$, is connected with a narrow current sheet located around the heliopause 
and extending into the heliotail.
As in the tail region the flow is directed along the field lines, the presence of
a current density immediately implies that even in such a field-aligned flow scenario
an electric field can exist due to the validity of resistive Ohm's law
$\vec E + \vec{\rm v} \times \vec B = \eta \vec j$, as long as the resistivity $\eta
\neq 0$. Furthermore, as was shown by \citet{2014arXiv1407.3227N}, the solution of 
non-ideal Ohm's law decouples from the rest of the MHD equations as the flow is 
field-aligned. Therefore, the only additional equation to be solved is 
\begin{equation}
\vec\nabla \times (\eta \vec j) = \vec 0\, ,
\end{equation} 
as $\vec E = \eta \vec j$ and $\vec\nabla \times \vec E = \vec 0$ (stationary 
approximation).

A reasonable resistivity should be valid in our steady-state model and 
account for the case of a collisionless plasma. 
While the Spitzer resistivity is effective only in collisional plasmas, the
turbulent collisonless resistivity (anomalous resistivity due to wave-particle interactions)
is usually not stationary. 

The usual approach for the resistivity is to use the electric force and to 
introduce some frictional force $\nu v_{D}$ acting on the charges $q$, with the 
collision frequency $\nu$ and the drift velocity $v_{D}$ \citep[e.g.][]{1977RvGSP..15..113P}
\begin{eqnarray} 
&&\frac{d v_{D}}{dt}=\frac{q}{m} E - \nu v_{D}=0 \quad\land\quad E=\eta j=\eta n q v_{D}\\ 
\Rightarrow && \eta=\frac{m\nu}{n q^2} \, .  
\end{eqnarray} 
For our collisionless plasma, we consider the interaction between the electromagnetic 
field and charged particles as a substitute for collisions. This ansatz is motivated by 
the fact that the interaction time, which is limited by the time the particle needs to 
cross the current sheet, i.e. the "transit time" of the particle within the system, is much 
shorter than the collision time. Hence the  collision time (collision frequency $\nu$) 
has to be replaced by the  gyro-time (gyro-frequency $q B/m$), delivering
\begin{equation}
\eta = \frac{1}{\sigma_{g}} \qquad \textrm{with} \qquad \sigma_{g}=\frac{n q}{|\vec B|}\, .
\end{equation}
where $\sigma_{g}$ is the gyroconductivity, $n$ and $q$ are the particle number density and 
charge, respectively.
This approach was introduced by \citet{1970P&SS...18..613S} and \citet{1985JGR....90.8543L}
and the resulting resistivity is called inertial or gyro-resistivity.  We want to emphasize that the 
substitution of the collision frequency by the gyro-frequency automatically delivers huge 
resistivity values in the case of a diluted plasma, which are only important for the generation 
of an electric field in regions of strong 
current density and not in regions where the field has the character of a potential field.

We use an approximate value for the magnetic shear of $B_{z}\approx 10^{-11}$\,Tesla, which is 
of the order of 10\% of the heliotail (or ISM) magnetic field strength and can be regarded as a
lower bound. For the typical lengthscale of the shear layer we set $l\approx 10^{3}$\,km 
\citep[e.g.,][]{1983MNRAS.205..839F}. Hence, we
can estimate the resulting current density via
\begin{equation}
|\vec j| \approx \frac{|B_{z}|}{\mu_{0} l} = 5.3\times 10^{-11}\,\textrm{A\,m}^{-2}\, .
\end{equation}
For density values of $10^{4}$\,m$^{-3}$ \citep{1986SSRv...43..329F} and magnetic field strengths
of $2\times 10^{-10}$\,Tesla typical for heliotail conditions, the gyroresisitivity $\eta$ is on 
the order of $10^{5}$\,Ohm\,m. Consequently, the parallel electric field becomes
\begin{equation}
    E_{\parallel} = \frac{\vec E\cdot \vec B}{|\vec B|} \approx
    \frac{\eta}{\mu_{0}}\,|\vec j| \approx 10^{-6}\,\textrm{Volt\,m}^{-1}\, .
\end{equation}

Within the tail, the field lines are stretched, and the current is concentrated 
around the heliopause region, providing a sufficiently extended
environment for accelerating particles continuously along the magnetic field 
lines. Outside this narrow region, the current vanishes and hence also the
electric field, so that those astrosphere regions can be ideal.

Considering a relatively conservative case, in which the field aligned electric 
field extends to about 100\,AU, only, meaning that the tail extends to just twice the distance than the heliopause nose, the voltage seen by the particles is
\begin{equation}
\int E_{\parallel} ds \approx E_{\parallel} \cdot s \approx 10^{7}\,\textrm{Volt}\, .
\end{equation}
This voltage can contribute to cosmic ray acceleration.

\section{Discussion and Conclusions}

We have shown that the distribution of null points and stagnation points 
defines the global topology and the large-scale structure of an astrosphere. 
Multiple separatrices can exist implying jumps (tangential discontinuities) of 
several physical parameters, such as the magnetic field strength, particle
density, etc. As the outermost separatrix defines the astropause,
its global geometrical shape is hence also determined.

With respect to the heliosphere, the multiple decreases and increases in the 
magnetic field strength as well as
in other physical parameters measured 
by {\it Voyager 1} \citep{2013Sci...341..147B} indicates several crossings of 
either one or several individual separatrices. Such a scenario is in good qualitative
agreement with the multiple separatrix structures due to more than one
null point as proposed here and formerly by \citet{2006A&A...454..797N}.
A similar scene considering multiple, nested separatrices and magnetic islands
was recently suggested based on detailed numerical simulations by
\citet{2013ApJ...774L...8S}.

Interestingly, our results for the two null point scenarios also agree with
the recently proposed presence of a heliocliff region inside the heliopause
\citep{2013ApJ...776...79F}. In particular, the heliocliff might be interpreted
as the separatrix resulting from the second null point (as shown in the middle
left panel of Fig.\,\ref{series}), and the streamlines originating from the
monopole part, which bend into the heliotail, would represent the open
heliosheath as introduced by \citet{2013ApJ...776...79F}. 
In the heliocliff region, the model of \citet{2013ApJ...776...79F} turns out
to produce a super-Alfv\'{e}nic field-aligned flow, while in our model
the flow close to the heliopause and in the heliotail region is 
field-aligned but can also be sub-Alfv\'{e}nic. 

Furthermore, the presence of magnetic shear flows can produce 
vortex current sheets \citep{2012AnGeo..30..545N} leading to the generation
of instabilities and magnetic reconnection close to separatrices. 
In the current work we restrict our analysis to a maximum of two
separatrices and we apply the mapping only to the heliotail with
one symmetric separatrix (top panel of Fig.\,\ref{axisym}) with two current
sheets.
As multiple separatrices can exist in the heliosphere, the presence of 
multiple curent sheets in the vicinity of these separatrices can lead to 
fragmented structures \citep[e.g.,][]{2013A&A...556A..61N, 2013ApJ...774L...8S}.
Introducing a non-collisional resistivity, strong electric (DC) fields parallel 
to the magnetic field can be generated, which can contribute to cosmic ray 
acceleration as suggested by \citet{2005PhDT........85N}.

\begin{acknowledgements}
We thank Andreas Kopp and two more, anonymous referees for helpful comments on
the paper draft.
This research made use of the NASA Astrophysics Data System (ADS). D.H.N. 
and M.K. acknowledge financial support from GA\,\v{C}R under grant numbers 13-24782S
and P209/12/0103, respectively. The Astronomical Institute Ond\v{r}ejov is supported by the
project RVO:67985815.
\end{acknowledgements}


\begin{thebibliography}{99}
\providecommand{\natexlab}[1]{#1}

\bibitem[{Arnol'd(1992)}]{Arnold}
Arnol'd, V.~I.: Ordinary Differential Equations, Springer, 1992.

\bibitem[{{Aschwanden}(2002)}]{2002SSRv..101....1A}
{Aschwanden}, M.~J.: {Particle acceleration and kinematics in solar flares - A
  Synthesis of Recent Observations and Theoretical Concepts},
  \ssr, 101, 1--227, 2002.

\bibitem[{{B{\'a}rta} et~al.(2010){B{\'a}rta}, {B{\"u}chner}, and
  {Karlick{\'y}}}]{2010AdSpR..45...10B}
{B{\'a}rta}, M., {B{\"u}chner}, J., and {Karlick{\'y}}, M.: {Multi-scale MHD
  approach to the current sheet filamentation in solar coronal reconnection},
  Advances in Space Research, 45, 10--17, 
  2010.

\bibitem[{{Bogoyavlenskij}(2000{\natexlab{a}})}]{2000JMP....41.2043B}
{Bogoyavlenskij}, O.~I.: {Counterexamples to Parker's theorem}, Journal of
  Mathematical Physics, 41, 2043--2057, 
  2000{\natexlab{a}}.

\bibitem[{{Bogoyavlenskij}(2000{\natexlab{b}})}]{2000PhRvE..62.8616B}
{Bogoyavlenskij}, O.~I.: {Helically symmetric astrophysical jets}, \pre, 62,
  8616--8627, 2000{\natexlab{b}}.

\bibitem[{{Bogoyavlenskij}(2001)}]{2001PhLA..291..256B}
{Bogoyavlenskij}, O.~I.: {Infinite symmetries of the ideal MHD equilibrium
  equations}, Physics Letters A, 291, 256--264, 2001.

\bibitem[{{Bogoyavlenskij}(2002)}]{2002PhRvE..66e6410B}
{Bogoyavlenskij}, O.~I.: {Symmetry transforms for ideal magnetohydrodynamics
  equilibria}, \pre, 66, 056410, 2002.

\bibitem[{{Burlaga} et~al.(2013){Burlaga}, {Ness}, and
  {Stone}}]{2013Sci...341..147B}
{Burlaga}, L.~F., {Ness}, N.~F., and {Stone}, E.~C.: {Magnetic Field
  Observations as Voyager 1 Entered the Heliosheath Depletion Region}, Science,
  341, 147--150, 2013.

\bibitem[{{Burlaga} and {Ness}(2014)}]{2014ApJ...784..146B}
{Burlaga}, L.~F. and {Ness}, N.~F.: {Voyager 1 Observations of the Interstellar
Magnetic Field and the Transition from the Heliosheath}, \apj, 784, 146, 2014.

\bibitem[{{Fahr} and {Neutsch}(1983a)}]{1983MNRAS.205..839F}
{Fahr}, H.~J. and {Neutsch}, W.: {Stationary plasma-field equilibrium states in
  astropause boundary layers. I - General theory}, \mnras, 205, 839--857, 1983.

\bibitem[{{Fahr} and {Neutsch}(1983b)}]{1983A&A...118...57F}
{Fahr}, H.~J. and {Neutsch}, W.: {Pressure distribution at the inner boundary of an astropause caused by a compressible stellar wind}, \aap, 118, 57--65, 1983.

\bibitem[{{Fahr} et~al.(1986){Fahr}, {Neutsch}, {Grzedzielski}, {Macek}, and
  {Ratkiewicz-Landowska}}]{1986SSRv...43..329F}
{Fahr}, H.~J., {Neutsch}, W., {Grzedzielski}, S., {Macek}, W., and
  {Ratkiewicz-Landowska}, R.: {Plasma transport across the heliopause}, \ssr,
  43, 329--381, 1986.

\bibitem[{{Fahr} et~al.(1993){Fahr}, {Fichtner}, and
  {Scherer}}]{1993A&A...277..249F}
{Fahr}, H.-J., {Fichtner}, H., and {Scherer}, K.: {Determination of the
  heliospheric shock and of the supersonic solar wind geometry by means of the
  interstellar wind parameters}, \aap, 277, 249, 1993.

\bibitem[{{Fichtner} et~al.(2014)}]{2014A&A...561A..74F}
{Fichtner}, H., {Scherer}, K., {Effenberger}, F., {Z{\"o}nnchen},
J., {Schwadron}, N., and {McComas}, D.~J.: 
{The IBEX ribbon as a signature of the inhomogeneity of the local
interstellar medium}, \aap, 561, A74, 2014.


\bibitem[{{Frisch} et~al.(2012)}]{2012ApJ...760..106F}
{Frisch}, P.~C., {Andersson}, B.-G., {Berdyugin}, A., 
	{Piirola}, V., {DeMajistre}, R., {Funsten}, H.~O., 
	{Magalhaes}, A.~M., {Seriacopi}, D.~B., {McComas}, D.~J., 
	{Schwadron}, N.~A., {Slavin}, J.~D. and {Wiktorowicz}, S.~J.:
{The Interstellar Magnetic Field Close to the Sun. II.},
\apj, 760, 106, 2012.

\bibitem[{{Fisk} and {Gloeckler}(2013)}]{2013ApJ...776...79F}
{Fisk}, L.~A. and {Gloeckler}, G.: {The Global Configuration of the Heliosheath
  Inferred from Recent Voyager 1 Observations}, \apj, 776, 79, 2013.

\bibitem[{{Gebhardt} and {Kiessling}(1992)}]{1992PhFlB...4.1689G}
{Gebhardt}, U. and {Kiessling}, M.: {The structure of ideal
  magnetohydrodynamics with incompressible steady flow}, Physics of Fluids B,
  4, 1689--1701, 1992.

\bibitem[{{Gurnett} et~al.(2013)}]{2013Sci...341.1489G}
{Gurnett}, D.~A., {Kurth}, W.~S., {Burlaga}, L.~F., and {Ness}, N.~F.:
{In Situ Observations of Interstellar Plasma with Voyager 1}, Science,
341, 1489--1492, 2013.

\bibitem[{{Hornig} and {Schindler}(1996)}]{1996PhPl....3..781H}
{Hornig}, G. and {Schindler}, K.: {Magnetic topology and the problem of its
  invariant definition}, Physics of Plasmas, 3, 781--791, 1996.

\bibitem[{{Karlick{\'y}} and
  {B{\'a}rta}(2008{\natexlab{a}})}]{2008CEAB...32...35K}
{Karlick{\'y}}, M. and {B{\'a}rta}, M.: {Fragmentation of Current Sheet},
  Central European Astrophysical Bulletin, 32, 35--38, 2008{\natexlab{a}}.

\bibitem[{{Karlick{\'y}} and
  {B{\'a}rta}(2008{\natexlab{b}})}]{2008SoPh..247..335K}
{Karlick{\'y}}, M. and {B{\'a}rta}, M.: {Fragmentation of the Current Sheet,
  Anomalous Resistivity, and Acceleration of Particles}, \solphys, 247,
  335--342, 2008{\natexlab{b}}.


\bibitem[{{Krimigis} et~al.(2013)}]{2013Sci...341..144K}
{Krimigis}, S.~M., {Decker}, R.~B., {Roelof}, E.~C., 
	{Hill}, M.~E., {Armstrong}, T.~P., {Gloeckler}, G., 
	{Hamilton}, D.~C. and {Lanzerotti}, L.~J.:
{Search for the Exit: Voyager 1 at Heliosphere's Border with the Galaxy}
Science, 341, 144-147, 2013.

\bibitem[{{Lau} and {Finn}(1990)}]{1990ApJ...350..672L}
{Lau}, Y.-T. and {Finn}, J.~M.: {Three-dimensional kinematic reconnection in
  the presence of field nulls and closed field lines}, \apj, 350, 672--691, 1990.

\bibitem[{{Lazarian} and {Desiati}(2010)}]{2010ApJ...722..188L}
{Lazarian}, A. and {Desiati}, P.: {Magnetic Reconnection as the Cause of Cosmic
  Ray Excess from the Heliospheric Tail}, \apj, 722, 188--196, 2010.

\bibitem[{{Lazarian} and {Opher}(2009)}]{2009ApJ...703....8L}
{Lazarian}, A. and {Opher}, M.: {A Model of Acceleration of Anomalous Cosmic
  Rays by Reconnection in the Heliosheath}, \apj, 703, 8--21, 2009.


\bibitem[{{Lyons} and {Speiser}(1985)}]{1985JGR....90.8543L}
{Lyons}, L.~R. and {Speiser}, T.~W.: {Ohm's law for a current sheet}, \jgr, 90,
  8543--8546, 1985.

\bibitem[{{McComas} et~al.(2013)}]{2013ApJ...771...77M}
{McComas}, D.~J., {Dayeh}, M.~A., {Funsten}, H.~O., {Livadiotis}, G. and {Schwadron}, N.~A.:
{The Heliotail Revealed by the Interstellar Boundary Explorer}, \apj, 771, 77, 2013.

\bibitem[{{Nerney} et~al.(1991){Nerney}, {Suess}, and
  {Schmahl}}]{1991A&A...250..556N}
{Nerney}, S., {Suess}, S.~T., and {Schmahl}, E.~J.: {Flow downstream of the
  heliospheric terminal shock - Magnetic field kinematics}, \aap, 250,
  556--564, 1991.

\bibitem[{{Nerney} et~al.(1993){Nerney}, {Suess}, and
  {Schmahl}}]{1993JGR....9815169N}
{Nerney}, S., {Suess}, S.~T., and {Schmahl}, E.~J.: {Flow downstream of the
  heliospheric terminal shock - The magnetic field on the heliopause}, \jgr,
  98, 15\,169, 1993.

\bibitem[{{Nerney} et~al.(1995){Nerney}, {Suess}, and
  {Schmahl}}]{1995JGR...100.3463N}
{Nerney}, S., {Suess}, S.~T., and {Schmahl}, E.~J.: {Flow downstream of the
  heliospheric terminal shock: Magnetic field line topology and solar cycle
  imprint}, \jgr, 100, 3463--3471, 1995.

\bibitem[{{Neutsch} and {Fahr}(1982)}]{1982MNRAS.202..735N}
{Neutsch}, W. and {Fahr}, H.~J.: {The magnetic and fluid environment of an
  ellipsoidal circumstellar plasma cavity}, \mnras, 202, 735--752, 1982.

\bibitem[{{Nickeler} and {Fahr}(2001)}]{2001ohnf.conf...57N}
{Nickeler}, D. and {Fahr}, H.~J.: {Stationary MHD-equilibria of the heliotail
  flow}, in: The Outer Heliosphere: The Next Frontiers, edited by {Scherer},
  K., {Fichtner}, H., {Fahr}, H.~J., and {Marsch}, E., p.~57, 2001.

\bibitem[{{Nickeler}(2005)}]{2005PhDT........85N}
{Nickeler}, D.~H.: {MHD equilibria of astrospheric flows}, Ph.D. thesis,
  Utrecht University, 2005.

\bibitem[{{Nickeler} and {Fahr}(2005)}]{2005AdSpR..35.2067N}
{Nickeler}, D.~H. and {Fahr}, H.-J.: {Reconnection at the heliopause}, Advances
  in Space Research, 35, 2067--2072, 2005.

\bibitem[{{Nickeler} and {Fahr}(2006)}]{2006AdSpR..37.1292N}
{Nickeler}, D.~H. and {Fahr}, H.-J.: {2D stationary resistive MHD flows:
  Borderline to magnetic reconnection solutions}, Advances in Space Research,
  37, 1292--1294, 2006.

\bibitem[{{Nickeler} and {Karlick{\'y}}(2006)}]{2006ASTRA...2...63N}
{Nickeler}, D.~H. and {Karlick{\'y}}, M.: {Are heliospheric flows magnetic
  line- or flux-conserving?}, Astrophysics and Space Sciences Transactions, 2,
  63--72, 2006.

\bibitem[{{Nickeler} and {Karlick{\'y}}(2008)}]{2008ASTRA...4....7N}
{Nickeler}, D.~H. and {Karlick{\'y}}, M.: {On the validity of ideal MHD in the
  vicinity of stagnation points in the heliosphere and other astrospheres},
  Astrophysics and Space Sciences Transactions, 4, 7--12, 2008.

\bibitem[{{Nickeler} and {Wiegelmann}(2010)}]{2010AnGeo..28.1523N}
{Nickeler}, D.~H. and {Wiegelmann}, T.: {Thin current sheets caused by plasma
  flow gradients in space and astrophysical plasma}, Annales Geophysicae, 28,
  1523--1532, 2010.

\bibitem[{{Nickeler} and {Wiegelmann}(2012)}]{2012AnGeo..30..545N}
{Nickeler}, D.~H. and {Wiegelmann}, T.: {Relation between current sheets and
  vortex sheets in stationary incompressible MHD}, Annales Geophysicae, 30,
  545--555, 2012.

\bibitem[{{Nickeler} et~al.(2006){Nickeler}, {Goedbloed}, and
  {Fahr}}]{2006A&A...454..797N}
{Nickeler}, D.~H., {Goedbloed}, J.~P., and {Fahr}, H.-J.: {Stationary
  field-aligned MHD flows at astropauses and in astrotails. Principles of a
  counterflow configuration between a stellar wind and its interstellar medium
  wind}, \aap, 454, 797--810, 2006.

\bibitem[{{Nickeler} et~al.(2013){Nickeler}, {Karlick{\'y}}, {Wiegelmann}, and
  {Kraus}}]{2013A&A...556A..61N}
{Nickeler}, D.~H., {Karlick{\'y}}, M., {Wiegelmann}, T., and {Kraus}, M.:
  {Fragmentation of electric currents in the solar corona by plasma flows},
  \aap, 556, A61, 2013.

\bibitem[{{Nickeler} et~al.(2014){Nickeler}, {Karlick{\'y}}, {Wiegelmann}, and
  {Kraus}}]{2014arXiv1407.3227N}
{Nickeler}, D.~H., {Karlick{\'y}}, M., {Wiegelmann}, T., and {Kraus}, M.:
  {Self-consistent stationary MHD shear flows in the solar atmosphere as electric field generators},
  \aap, in press (arXiv:1407.3227).

\bibitem[{{}(1977)}]{1977RvGSP..15..113P}
{Papadopoulos}, K.: {A review of anomalous resistivity for the ionosphere},
Reviews of Geophysics and Space Physics, 15, 113-127, 1977.

\bibitem[{{Parker}(1961)}]{1961ApJ...134...20P}
{Parker}, E.~N.: {The Stellar-Wind Regions.}, \apj, 134, 20, 1961.

\bibitem[{{Parnell} et~al.(1996){Parnell}, {Smith}, {Neukirch}, and
  {Priest}}]{1996PhPl....3..759P}
{Parnell}, C.~E., {Smith}, J.~M., {Neukirch}, T., and {Priest}, E.~R.: {The
  structure of three-dimensional magnetic neutral points}, Physics of Plasmas,
  3, 759--770, 1996.

\bibitem[{{Sahai} and {Chronopoulos}(2010)}]{2010ApJ...711L..53S}
{Sahai}, R. and {Chronopoulos}, C.~K.: {The Astrosphere of the Asymptotic 
Giant Branch Star IRC+10216}, \apjl, 711, L53-L56, 2010.

\bibitem[{{Scherer} and {Fichtner}(2014)}]{2014ApJ...782...25S} 
{Scherer}, K. and {Fichtner}, H.: {The Return of the Bow Shock},
\apj, 782, 25, 2014.

\bibitem[{{Speiser}(1970)}]{1970P&SS...18..613S}
{Speiser}, T.~W.: {Conductivity without collisions or noise}, \planss, 18, 613, 
  1970.

\bibitem[{{Suess} and {Nerney}(1990)}]{1990JGR....95.6403S}
{Suess}, S.~T. and {Nerney}, S.: {Flow downstream of the heliospheric terminal
  shock. I - Irrotational flow}, \jgr, 95, 6403--6412, 1990.

\bibitem[{{Swisdak} et~al.(2013){Swisdak}, {Drake}, and
  {Opher}}]{2013ApJ...774L...8S}
{Swisdak}, M., {Drake}, J.~F., and {Opher}, M.: {A Porous, Layered Heliopause},
  \apjl, 774, L8, 2013.

\bibitem[{{Ueta}(2008)}]{2008ApJ...687L..33U}
{Ueta}, T.: {Cometary Astropause of Mira Revealed in the Far-Infrared},
\apjl, 687, L33-L36, 2008.

\end{thebibliography}


\end{document}